\documentstyle{mn}
\input{psfig}

\begin{document}

\newcommand{\HI}{H{\,\small I}}
\newcommand{\kms}{km s$^{-1}$}

\title[Centaurus A: multiple outbursts or bursting bubble?]{Centaurus A:
multiple outbursts or bursting bubble?}

\author[R. Morganti et al.]{R.
Morganti$^{1,2,}$\thanks{Email:rmorgant@ira.bo.cnr.it}, N.E.B. Killeen$^2$,
R.D. Ekers$^2$, T.A. Oosterloo$^{2,3}$
\\
$^1$ Istituto di Radioastronomia, CNR, via Gobetti 101, 40129 Bologna,
Italy \\
$^2$ Australia Telescope National Facility, CSIRO, PO Box 76,
Epping, NSW\,2121, Australia \\
$^3$ Istituto di Fisica Cosmica,
CNR, via Bassini 15, 20133 Milan, Italy}

\date{Accepted~~, Received~~}

\maketitle
\begin{abstract}

We present new radio observations of the brighter region of the northern lobe
(the Northern Middle Lobe, NML) of Centaurus A obtained at 20 cm with the
Australia Telescope Compact Array.  The angular resolutions are $\sim$50 and
$\sim$130 arcsec, therefore much higher than for the previously available
radio images of this region.  The most interesting feature detected in our
images is a {\sl large-scale jet} that connects the inner radio lobe and the
NML and that is imaged for the first time.  The NML itself appears as diffuse
emission with a relatively bright ridge on the eastern side.  

The radio morphology of Centaurus~A and, in particular, its NML could be the
result of a precessing jet that has undergone a strong interaction with the
environment at least in the northern side.  The very big drop in intensity
between the inner jet and the large-scale jet can be explained with a sequence
of bursts of activity at different epochs in the life of the source. 
Alternatively (or additionally) a ``bursting bubble'' model is proposed to
explain this big drop in intensity, which could also explain the good
collimation of the large-scale jet.  In this model, the plasma accumulated in
the inner lobe would be able to ``burst'' out only through one nozzle that
would be the region where the large-scale jet forms.  The location of the
nozzle would represent a region where the pressure gradient is more
favourable.

From the comparison between the radio emission and the regions of ionized gas
discovered by Graham \& Price (the so-called optical filaments) we find that
the inner optical filament ($\sim 8$ kpc from the centre) falls about 2 arcmin
($\sim 2$ kpc) away from the large-scale radio jet.  Thus, the inner filament
does not seem to have experienced a {\sl direct} interaction with the radio
plasma.  The complex velocity field observed in this filament could be
therefore due to strong instabilities produced by the ``bursting bubble''. 
The outer filaments appear to be, in projection, closer to and aligned with
the radio emission in the transition region between the jet and the lobe,
arguing for a direct interaction with the radio jet.  However, also in this
case a more complicated interaction than assumed so far has to be occuring
because of the relative position of the ionized and neutral gas regions
compared to the radio jet as well as the kinematics of the ionized gas.

\end{abstract}

\section {Introduction} 

There are a number of phenomena observed in radio galaxies that require a
good knowledge of the radio morphology (on different scales) in order to be
better understood.  For example, the effects of the external medium can be
revealed in the way they modify/disrupt the radio morphology.  Sharp
changes and bends in the morphology can be used as tracer of the external
conditions.  Furthermore, the radio morphology can be related to the way the
plasma is transported from the active nucleus to the outer lobes.  The
relation between the large scale morphology (Fanaroff-Riley (FR) type,
Fanaroff \& Riley 1974) of the sources and, e.g., the characteristics of the
jets in radio galaxies is well-known (Bridle \& Perley 1984, Laing 1993).  For
example, in the case of FR class I sources, the observed large scale radio
morphology has been used as indication that the jets are turbulent and low
Mach-number on large scale with entrainment playing an important role in
shaping their morphology (Bicknell 1986, Bicknell et al.  1990).  Finally, the
interstellar and/or the intergalactic medium (ISM/IGM) around radio sources
can be compressed and perhaps also ionized by the interaction with the radio
plasma.  

As the nearest radio galaxy, Centaurus~A (NGC~5128) has been very often used as
a ``laboratory'' object where to study some of the phenomena that are typical
of active galactic nuclei and that only in this object can be studied in
detail.  A comprehensive review of the present state of knowledge of
Centaurus~A is given by Israel (1998).  However, for studying the radio
emission, the vicinity of Centaurus~A has often been a problem for the
observations.  Although Centaurus~A is a radio source of modest power ($ P_{\rm
2.7 GHz} = 10^{24.26}$ W Hz$^{-1}$), which places it among the low-luminosity
radio galaxies typically classified as FR~I, its radio flux is very high
($S_{\rm 2.7 GHz}=128$ Jy, Wright \& Otrupcek 1990).  Moreover, the entire radio
structure of Centaurus~A covers an area of about $10^{\circ} \times 5^{\circ}$
on the sky\footnote{in fact, Cen A is one of the very few extragalactic objects
in the sky for which perspective has to be taken into account!}.  This huge
angular size (that corresponds to a projected linear size\footnote{we assume a
distance to Centaurus A of 3.7 Mpc (Hui et al.\ 1993).  This gives 1 arcsec =
18 pc, 1 arcmin $\simeq$ 1 kpc.} of $\sim$650 kpc) and the strong emission from
the central regions makes it very difficult to carry out observations at
intermediate resolution of the large scale structure.  This has limited the
radio observations to either low resolution images or to detailed studies of
the central region, as can be seen in Fig.~1 where some of the available radio
images are collected.  Thus, although Centaurus~A would be an ideal object for
studying in detail the radio morphology and its relation with the surrounding
medium, its vicinity makes this task quite difficult. 

So far, its large-scale radio structure has been imaged only using single-dish
Parkes observations with resolutions ranging from a few to several arcminutes
(Cooper, Price \& Cole 1965, Junkes et al.\ 1993).  An image of the whole
structure of the radio emission at 4.7 GHz (taken from Junkes et al.\ 1993) is
shown in Fig.~1a,b.  The dominant structures at these resolutions are the
low-brightness very extended lobes (the Outer Lobes).  A bright central region
(the Inner Lobes) is visible in the Parkes radio image and it includes (at 21
cm) $\sim 22$\% of the observed total flux (Cooper et al.\ 1965).  This central
region has been studied at high resolution with the VLA by Burns et al.\ (1983)
and Clarke et al.\ (1992) and an image is shown in Fig.~1c.  On the VLBI
(sub-parsec) scale a jet and a counter-jet have been detected in the same
position angle as the arcsec scale jet (Fig.~1d, Jones et al.  1996).

The two Outer Lobes are not completely symmetrical.  The Northern Outer Lobe
includes, at about 30 arcmin ($\sim 30$ kpc) from the centre, the so-called
Northern Middle Lobe (NML), which is the brightest region in the Northern Outer
Lobe and does not have a counterpart in the southern lobe. 

The region of the NML is particularly interesting because of a number of
features observed there:  the {\sl extended emission line regions  of
highly ionized gas} (also known as {\sl ``the optical filaments''}) discovered
by Blanco et al.\ (1975) and situated in the region between the Northern Inner
Lobe and the NML (at distances ranging from $\sim$7.5 arcmin up to 28 arcmin
from the nucleus); the {\sl soft X-ray emission} associated with the NML and
detected with the {\sl Einstein Observatory} (Feigelson et al.\ 1981); a cloud
of {\sl neutral hydrogen} detected by Schiminovich et al.\ (1994) next to the
outer optical filament.  These features are {\sl only observed in and around
the NML}, and together with the fact that there is no counterpart of the NML in
the Southern Outer Lobe, they indicate that in this region of the radio source
some kind of interaction may be happening between the radio plasma and the
ISM/IGM of Centaurus~A. 

The optical filaments are particularly interesting.  They are characterized by
a very high ionization level and their spectra have been investigated in
detail by Morganti et al.\ (1991, 1992).  Two possible explanations have been
suggested to account for their high ionization.  One possibility is that the
gas is photoionized by UV radiation from the nucleus (Morganti et al.\ 1991). 
The other possibility is that they are the result of a strong interaction
between the radio plasma and the ISM (Sutherland et al.\ 1993).  The latter
explanation would require a degree of correspondence between the ionized gas
of the filaments and the radio structure.  The radio/optical interaction has
been also claimed to be the most likely mechanism able to explain the complex
velocity fields observed in the filaments.  
A number of objects have been studied so far (see
e.g.  PKS~2250-41, Clark et al.\ 1997 and Villar-Martin et al.\ 1998a;
PKS~1934-46, Villar-Martin et al.\ 1998b; 3C171, Clark et al.\ 1998;
PKS~2152-69, Fosbury et al.\ 1998) and Centaurus~A can be an ideal object for
a study of this kind.

\begin{figure*}
%\centerline{\psfig{figure=F1-MY981.ps,width=12cm,angle=0}} 
\caption{Collection of some of the available radio images of Centaurus A: a)
large scale structure from 6cm Parkes observations (Junkes et al. 1993;
vectors show the position angle of the electric field and their length is proportional to the
polarized intensity); b) an
enlargement of the northern lobe (Junkes et al. 1993); c) 6cm VLA image of the
inner lobes (Burns et al. 1983); d) VLBI jet from Jones et al. (1996). The
coordinates are B1950. }
\end{figure*}

It follows from the above that, in order to investigate the effects of the
interaction between the radio plasma and the environment, the evolution of the
jet and the connection (if any) between the inner radio jet and the NML, as well
as the effect of the radio plasma on the optical filaments, radio observations
of this region at relatively high resolution may give valuable information. 

Here we present the results of new observations done with the Australia
Telescope Compact Array (ATCA) at 20 cm of the region of the NML.  The main
goal of these observations is to investigate the structure of this lobe at
higher resolution compared to the previous single-dish images, to investigate
the polarization in this lobe and to study the radio morphology in connection
with the optical filaments and other structures present in the region of the
NML.

\section {ATCA observations }

The observations were done on 5, 11 and 23 June 1995 using three of the four
standard 750-m configurations available with ATCA (the data obtained with the
fourth 750-m configuration (750A) were of much poorer quality and were
therefore not used).  All the configurations were combined together in order
to get a good {\it uv} coverage.  Observations with the 375-m configuration
were carried out in March 1998. 
In all the observations the central frequency
was 1392~MHz and the standard 32~channels, 128~MHz bandwidth
continuum mode was used.

%
%%% Table 1
%
\begin{table*}
\begin{center}
{\sc Table 1.} {Log of the observations}
\smallskip

\begin{tabular}{ccc|ccccc} \hline\hline
 Fields  & \multicolumn{2}{c}{Pointing Centres}  & $\nu$ &  Configurations & Min-Max Baseline & 
rms$_{\rm I}$ & rms$_{\rm P}$   \\
         &    R.A.       &        Dec.      &           &                 &   meters 
&  \multispan{2} {mJy/beam}   \\ \hline
  \#1  & 13:25:24 & -43:01:11     & 1343   & 750B,C,D  & 31-765 & 15.3 & ... \\
  \#1  & 13:25:24 & -43:01:11     & 1384   & 375       & 31-459 & 18.8 & 11.5 \\
  \#2 & 13:26:54    & -42:32:36   & 1343   & 750B,C,D  & 31-765 & 6.7 & 2.4 \\
  \#2 & 13:26:54    & -42:32:36   & 1384   & 375       & 31-459 & 9.7 & 3.0 \\
\hline\hline
\end{tabular}
\end{center}
\end{table*}

\begin{figure*} 
\centerline{\psfig{figure=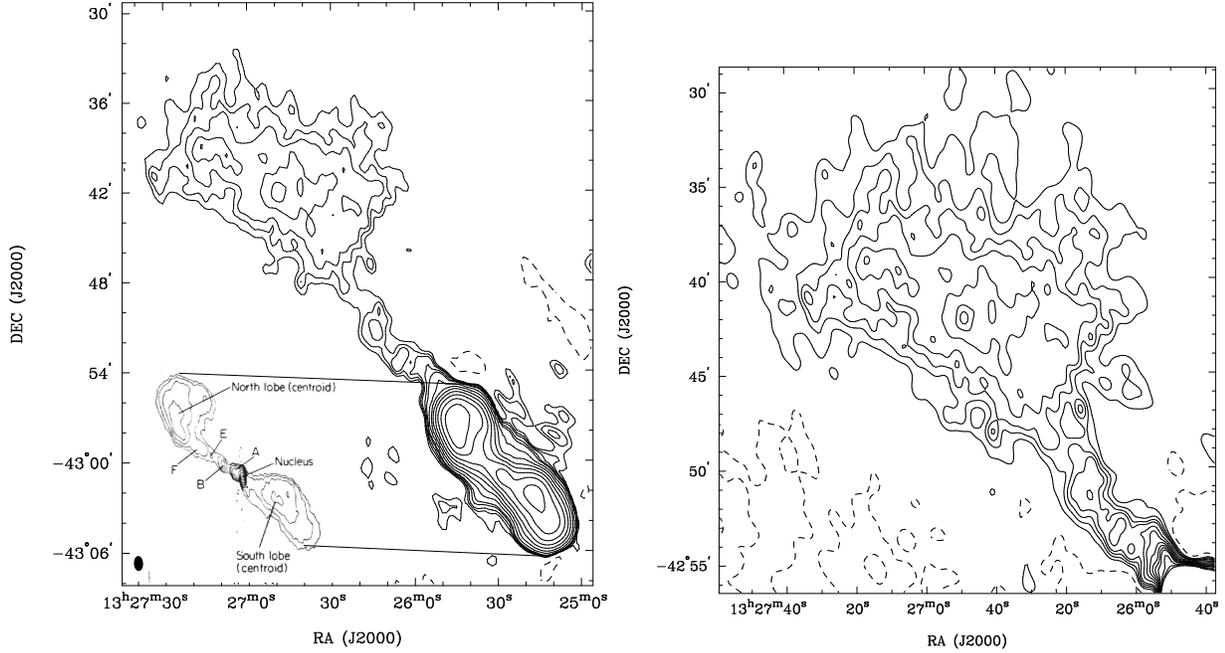,width=8cm,angle=0}
\psfig{figure=F2b-MY981.ps,width=8cm,angle=-90}}
\caption{{\sl Left:} Image of the NML obtained with the combined data 
from the 750-m
configuration.  The contour levels are $0.025 \times
-3, 1, 1.5, 2, 3, 4, 8, 16, 32, 64, 128, 256, 512, 1024$ Jy beam$^{-1}$.  
The inset shows the
higher resolution VLA image from Burns et al.  (1983). {\sl Right:} an image
of field \#2 that shows in more detail the radio structure. Given the lower
noise of this field compared to the central one (see Table~1), the contour levels are lower
that in the combined image on the left. The contour levels are $0.0125 \times
-2, 1, 2, 3, 4, 6, 8, 10, 12, 14, 16, 20, 24, 28$ Jy beam$^{-1}$.
 } 
\end{figure*}

\begin{figure}
\centerline{\psfig{figure=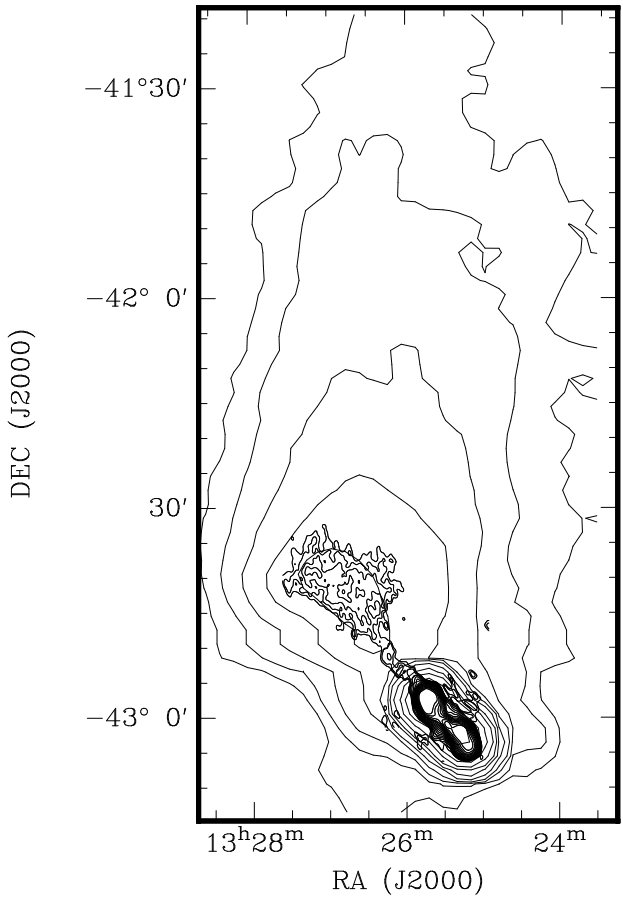,width=10cm}}
\caption{The same image as in Fig.2 superimposed to the Parkes 6cm image
(Junkes et al. 1993) for a
more direct comparison of the structure with the increase in resolution.}
\end{figure}

In this paper we will present and discuss the results obtained for two pointing
centres: the central field (i.e.  the field centred on the core of Centaurus A,
defined as field \#1 in Table 1) and the field centred north of the NML
(defined as field \#2 in Table 1).  The latter turned out to be the best for
mapping the NML and this is likely due to the attenuation of the central very
bright region by the primary beam. 

The source was observed for 12h in each configuration.
The data were calibrated by using the MIRIAD package (Sault, Teuben \&
Wright 1995), which is necessary for the polarization  calibration of
ATCA data.  The flux scale is based on the recent compilation of measurements
of the primary calibrator PKS~1934-638 by Reynolds (1996) which corresponds to
14.9 at 1.4~GHz.  We used PKS~1315-46 as a secondary calibrator with a flux of
2.20 Jy at 1.4~GHz. 

The images were produced using uniform weighting.  The resolution of the image
obtained combining the data from the 750-m configurations is $56 \times 36$
arcsec (in p.a.=2$^\circ$) while it is $135 \times 75$ arcsec (in
p.a.=--3$^\circ$) for the images obtained using the data from the 375-m
configuration.  These resolutions correspond to an average linear scale of
$\sim 0.8$ kpc and $\sim 1.9$ kpc respectively.  The rms noise of the $I,Q$
and $U$ maps varies for the two pointing centres and they are listed in
Table~1.

Together with the total intensity images we have also obtained the images for
the Stokes parameters ($Q, U$), the polarized intensity 
($P=(Q^2+U^2)^{1/2}$) and position-angle ($\chi= 0.5 {\arctan} (U/Q)$) images. 
The polarized intensity and the fractional polarization ($m=P/I$) were
estimated only for the pixels for which $P>5\sigma_{QU}$. 

In the case of the data from the 375-m configuration, we have estimated the
polarization for the channels at the edges and in the middle of the band (1372,
1416~MHz and 1392~MHz).  This has allowed us to derive the Faraday rotation
(although we are somewhat limited by the small difference between the
frequencies).

\section {Results}

\subsection {Total intensity}

Fig.~2 left shows the image (corrected for the primary beam) obtained using
the 750-m configuration data (i.e.  our higher resolution data) and resulting
from linearly mosaicing the images from the two pointing centres, and Fig.~2
right shows an image of field \#2.  Fig.~3 shows the same image superimposed
to the Parkes image to allow a more direct comparison of the morphology at
different resolutions. 
The most interesting feature detected in our new images is the {\sl jet-like}
structure that connects the inner lobe and the NML.  Although the nature of
this structure is not yet fully understood (see \S 4.1), throughout the paper
we will refer to it as the {\sl large-scale jet} (to distinguish it from the
inner jet imaged by the VLA).  The large-scale jet 
was not known before and it extends from 6
arcmin up to 15 arcmin (i.e.  from $\sim 6$ to about 15 kpc) from the centre. 
The large-scale jet is not perfectly straight but ``wiggles'' slightly.  At
larger distances from the nucleus, this jet widens to form the NML.  The
large-scale jet has a position angle of $\sim 45^\circ$ and is therefore
slightly misaligned compared to the inner jet that has a position angle $\sim
55^\circ$, as observed by the VLBI (Jones et al.  1996, Tingay et al.  1998)
and by the VLA (Clarke et al.  1992). 
  
The NML appears as diffuse emission with a relatively bright ridge and a sharp
edge on the eastern side, more evident in the low resolution image.  This
bright ridge appears to represent the continuation of the large-scale jet
(suggested also by the agreement in the position angle, both at $\sim
45^\circ$).  The NML is the place where another change in the position angle
of the radio structure occurs from $\sim 45^\circ$ to $\sim -10^\circ$
corresponding to the position angle of the northern Outer Lobe (Cooper et al. 
1965, Junkes et al.  1993; see Fig.~1).  Our lower resolution image (from the
375-m configuration data; see Fig.~4)
shows part of the very low brightness region of the NML that extends in the NW
direction.  

Finally, we note that the peak of the NML is only 365.6 mJy beam$^{-1}$ (at 1392~MHz) in
the image from the 375-m data.  Compared with the peak of the inner lobe from
the same data, 56.1 Jy, this shows the considerably lower brightness of the
NML (ratio 1:157) and, therefore, the difficulty of imaging the extended
emission in the NML at this intermediate resolution. 

\begin{figure}
\centerline{\psfig{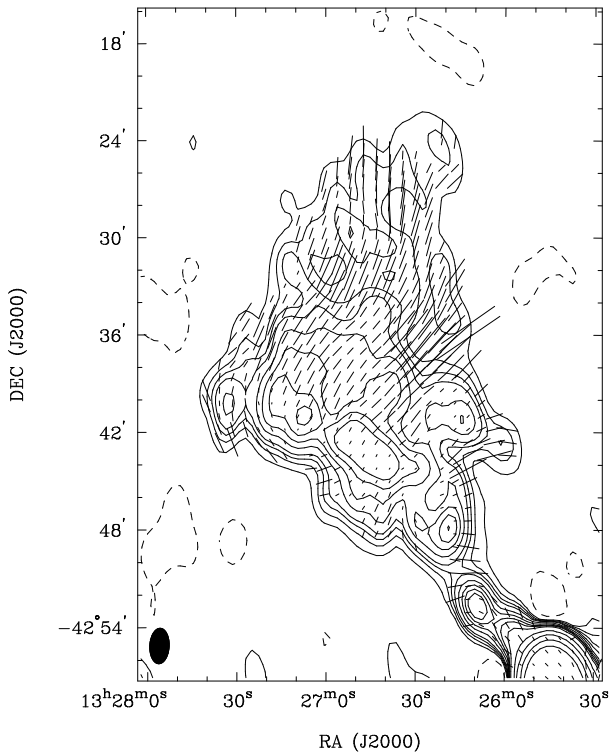}} \caption{Image
of the NML obtained with the combined data from the 375-m configuration with
superimposed vectors indicating the projected magnetic field direction.
The vectors are proportional in length to the fractional polarization ($m$)
with 10$^{\prime\prime}$ corresponding to 20\%.  The contour levels are:
$0.030 \times -2.5,1,2,3,4,6,8,10,12,16,32,64,128,256$ Jy beam$^{-1}$.  The
magnetic field has been corrected for an average $RM = -60$ rad m$^{-2}$.}

\end{figure}

\subsection {Polarization}

The analysis of the polarization presented here has been done using the data
from the 375-m configuration (similar results have also been obtained with the
750-m data and we omit these for brevity).  Fig.~4 shows the contours of the
total intensity for field \#2 from the 375-m configuration data, with
superimposed vectors whose length is proportional to the fractional
polarization and whose position angle is that of the magnetic field corrected
for an average $RM = -60$ rad m$^{-2}$ (see below).  The polarization reveals a number of
interesting structures that were not evident in the low resolution Parkes
images. 

The large scale jet appears to be clearly polarized with a position angle of
the magnetic vectors parallel to the jet in the region where it connects to
the inner lobe.

Along the large-scale jet, at about $\sim 12$ kpc from the nucleus, the
magnetic vectors turn perpendicular to the direction of the jet.  In polarized
emission the large-scale jet seems to bend west instead of proceeding to the
brighter part (ridge) of the NML.  The brighter ridge in the NML shows indeed
a lower fractional polarization with a ``corridor'' of low polarization or
even unpolarized emission in the east side of the lobe.  The average
fractional polarization in the bright ridge is between 10 and 20 \%.  On the
contrary, the western region of the lobe, shows  very high polarization
(between 50 and 60 \%) consistent with the fact that the position angles of
the magnetic vectors are very constant across this region 
with a typical p.a.  of $\sim 140^\circ$, indicating that no
major inhomogeneities are present on the scale of our observations.  The high
fractional polarization and the position angle of the magnetic vectors are in
agreement with previous observations (Cooper et al.  1965, Junkes et al.  1993
from 5-GHz data, see Fig.  1b) despite the large difference in resolution. 
This indicates that the effects of depolarization due to the beam are not very
strong.
%{\tt check paper Laing about high degree of polarization even if the field is
%not ordered}

We have used the images made with the channels at the edges and in the middle
of the band (1372, 1416 MHz and 1392 MHz) to estimate the rotation measure
($RM$).  The use of three channels in the band allows us to avoid $n\pi$
ambiguity in the estimate of the $RM$.  The $RM$ is defined as
$\chi(\lambda^2) = \alpha + RM\lambda^2$, where $\alpha$ is the intrinsic
position angle and $\chi$ the apparent position angle at the $\lambda$ of the
observations.  Fig.~5 shows the obtained rotation measure.  Large scale
variations are present although one should keep in mind that in some areas the
errors on the derived values are quite large, up to $\sim 40$ rad m$^{-2}$. 
The average rotation measure is $\sim -60 \pm 10$ rad m$^{-2}$.  This is in
agreement with what found by Cooper et al.  (1965; see also Junkes et al. 
1993).  The position angle of the magnetic vectors has been corrected for this
average value. 

In order to investigate how reliable the RM measurements are, we have done a
similar study of the polarization (i.e.  using the channels at the edges of
the band) for the central field.  Fig.~6 shows the total intensity for field
\#1 with superimposed vectors whose length is proportional to the fractional
polarization and whose position angle is that of the magnetic field (for a
frequency of 1392~MHz).  The average fractional polarization in the northern
inner lobe is between 10 and 25\% while the rotation measure is between --45
and --65 rad m$^{-2}$.  These values can be compared with what obtained from
the previous VLA studies (Clarke et al.  1992 and Clarke priv.  communication)
although the large difference in resolution has to be taken into account. 
Clarke et al.  (1992), from their arcsec resolution images, found a fractional
polarization in the northern inner lobe between 11 and 25 \% for the brighter
regions, going up to $\sim 40 -50$\% for the more diffuse low brightness
parts.  The RM they found (again in the northern inner lobe) is between --43
and --65 rad m$^{-2}$.  Therefore our values are in very good agreement with
those from the VLA, despite the large difference in resolution. 

\subsection{Radio and X-ray emission}

It is useful, for the discussion that will follow, to compare the new radio
image with the X-ray emission in the NML.  An association between soft X-ray
and the radio emission in the NML was detected with the {\sl Einstein Observatory} by
Feigelson et al.  (1981).  A look at the ASCA GIS low-energy (0.7 -- 2 keV)
image (pointed on the NML) from the public archive appears to confirm the
extended soft X-ray emission in the region of the NML and aligned with the
brighter radio ridge on the eastern side.

Feigelson et al.\ (1981) discuss three possibilities for the origin of the
X-ray emission: inverse Compton scattering of the microwave background,
synchrotron emission and thermal processes.  Inverse Compton scattering was
ruled out by Feigelson et al., because the required magnetic field (0.2
$\mu$G) was not consistent with the value derived from the old radio data
using the standard equipartition arguments.  Following the formulae in
Pacholczyk (1970) we have estimated the minimum magnetic field in the case of
equipartition: $ H_{\rm eq} \propto [(1 + k) \phi^{-1} \theta^{-3} S_\nu
D]^{2/7}$ where $k$ is the ratio between the energy of the heavy particles and
of the electrons, $\phi$ is the filling factor, $\theta$ the source size and
$D$ the distance.  With the parameters derived from our new data, we confirm
that a magnetic field of about 2 $\mu$G characterizes the NML, therefore too
high for Inverse Compton scattering.  This value has been derived using the
commonly used value $k=1$ (and $\phi = 1$) and the situation is, of course,
even worse if we assume an higher value for $k$ (see discussion in Killeen et
al.  1988).  In other words, based on the equipartition magnetic field, we
would expect an higher X-ray flux. 

When our radio image is compared with the soft X-ray image from ASCA, the
X-ray emission appears co-spatial with the brighter ridge of radio emission in
the eastern side of the NML.  However, the X-ray emission is weak ($\sim 1
\times 10^{39}$ ergs s$^{-1}$ in the range 0.5 -- 4.5 keV according to
Feigelson et al.  1981) and it appears to be very difficult to investigate a
possible coincidence between the X-ray and radio peaks (that could support the
synchrotron hypothesis, but see also Mack, Kerp \& Klein 1997).  The X-ray appears more diffuse and especially in the
south-west part not coincident with the peak of the radio emission.  However,
a more detailed reduction of the ASCA X-ray data is necessary to confirm this. 

According to Feigelson et al.\ (1981) the most likely explanation for the soft
X-ray emission in the region of the NML, are thermal processes produced by
compression of the interstellar medium by the radio lobe.  Interestingly, the
X-ray emission appears to be associated with the region of lower fractional
polarization in the NML.  This coincidence indeed suggests that this is the
process responsible for the X-ray emission and that the X-ray
gas is producing a depolarizing screen around the radio lobe.  A study of the
depolarization will allow to investigate this hypothesis. 

\begin{figure}
\centerline{\psfig{figure=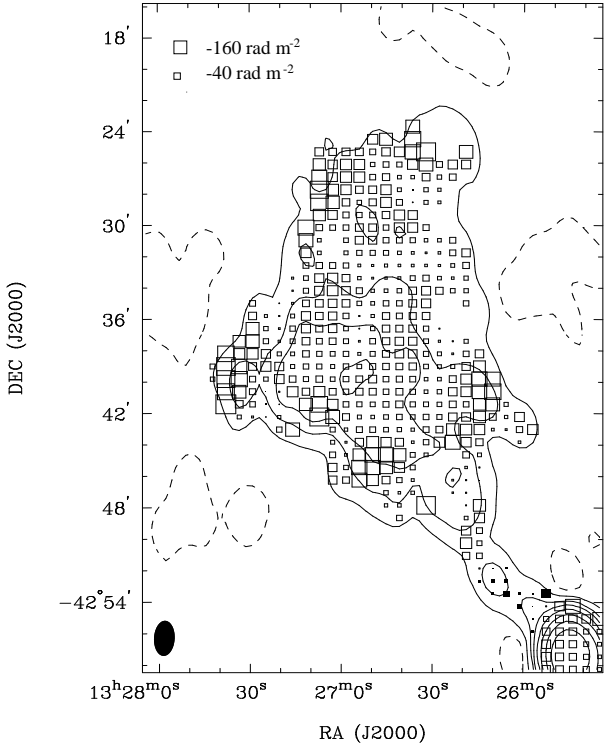,width=8cm,angle=0}} 
\caption{Image of the NML obtained with the combined data from the 375-m
configuration with superimposed boxes proportional to the rotation 
measure ($RM$). Filled squares represent positive numbers for the $RM$ while
empty squares represent negative numbers (the overall range of values 
is between $\sim +40$ and $\sim -160$ rad m$^{-1}$).}
\end{figure}

\begin{figure}
\centerline{\psfig{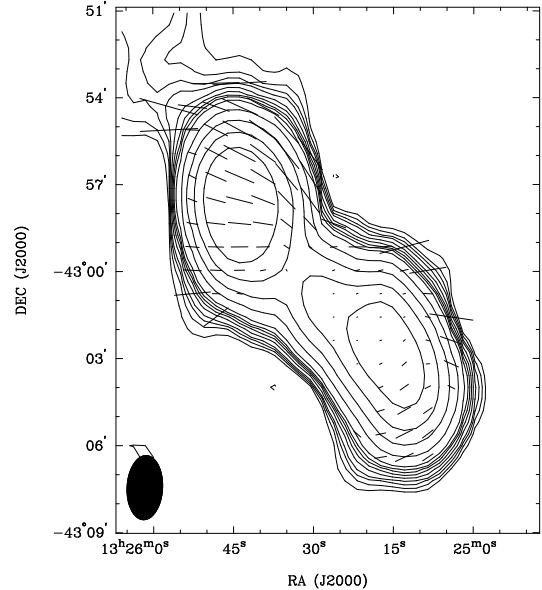}} 
\caption{ Image of the inner lobe from the 375-m 
configuration with
superimposed vectors indicating the projected magnetic field direction.
The vectors are proportional in length to the fractional
polarization ($m$) with 10$^{\prime\prime}$ corresponding to 7\%. 
The contour levels are $0.08 \times
-3, 1, 2, 3, 4, 6, 8, 10, 12, 16, 32, 64, 128, 256$ Jy beam$^{-1}$.}
\end{figure}

\section{Discussion}

\subsection{The nature of the NML}

The new radio images allow us to investigate in more detail the structure of
the NML in Centaurus~A.  We shall now attempt to understand how the large
scale radio structure, and in particular the large-scale jet revealed by our
new images, can be explained. 

\subsubsection{Interaction, change of axis and multiple outbursts}

In Centaurus A, the precession of the central engine, as described by Haynes
et al.  (1983), is strongly suggested by the overall {\bf S}-shape of the
large-scale radio structure.  However, some of the observed characteristics
have been also described as the result of interaction between the radio plasma
and the ISM/IGM.  In particular, the abrupt change in the observed position
angle in coincidence of the NML, as well as the asymmetry in the radio
structure (i.e. the absence of a middle lobe in the southern part), would not be easy to reconcile only with a smooth precession of the
engine.  Moreover, interaction is the most likely explanation for the
association between the radio and the soft X-ray emission detected in the NML
(Feigelson et al.  1981, as already discussed in \S 3.3).  Thus, the radio
structure of Centaurus~A can be described as the result of a jet that has
changed its position angle during its life and that has undergone
also a strong interaction with the environment at least in the northern side. 
It is also worth mentioning that Centaurus~A is believed to be the result
of a ``recent'' merger from where the neutral gas and the dust lane have been
accreted.  Such a merger may have disturbed the position angle of the
radio jet as a result of change of axis of the inner disk (Schreier et al. 
1998; Pringle 1997).

Similarities between Centaurus A and M87 (Virgo~A) have been already pointed
out by other authors and therefore it is not surprising to see that similar
hypotheses (in particular those described in \S 4.1.1) have been suggested to
explain the radio morphology of these two radio galaxies.  Indeed, there seems
to be a number of analogies between Centaurus A and M87.  In particular, as in
M87 (see Klein 1998 for a review), the structure of Centaurus~A can be
separated in three components.  The three components (inner, middle and outer
lobes) show in both cases an abrupt change in position angle.  To describe the
various structures observed in M87, Feigelson et al.\ (1981) proposed either
the ejection of inhomogeneous jets by a precessing nucleus or the continuous
ejection of jets that bend because of  interactions with the environment. 
Klein (1998) proposed, as the simplest interpretation, a change of orientation of
the ejection during different epochs of activity.  In the case of M87, in
fact, the structure seems to be difficult to reconcile with being powered by a
jet that had constant thrust and orientation over the past 10$^8$ years. 

On the other hand, M87 and Centaurus~A have also a number of differences. 
Apart from the radio power, the main difference is in the environment: Virgo~A
is part of a rich cluster and this is more likely to have a strong influence
on the morphology of the radio emission.  Indeed, the bend in the structure of
Centaurus~A is not as sharp as observed in M87, probably indicating a less
strong interaction.

However, the scenario mentioned above does not explain the large drop in the
brightness that is observed between the inner and the outer lobe region.  One
way of explaining this characteristic is assuming multiple outbursts due to
different epochs of activity in the life of the source.  The inner lobe would
be the result of the most recent (and still active) outburst, while the
large-scale jet and the outer lobe would then represent relics of past
activity.  The high fractional polarization and ordered magnetic field in the
NML would be consistent with this hypothesis.  Polarization studies of tailed
radio galaxies (e.g., Feretti et al.\ 1998a) show that the degree of
polarization increases along the tails, reaching often values up to (and
sometimes larger than) 50\%, therefore indicating a very ordered component of
the magnetic field.  The same must occur in the more relaxed part of the
Centaurus~A lobe.

\subsubsection{An alternative scenario: a ``bursting bubble''}

The need for an alternative model or an additional ingredient is due to the
fact that the large-scale jet appears to be quite collimated, more than what
would be expected in a relic-like kind of structure.  The alternative model
comes from the analogy between Centaurus A and Wide Angle Tails (WAT) radio
galaxies.  Some of these objects show sharp bends and jet disruption followed
by a recollimation of the jet itself that can propagate for a long distance
downstream the deflection point.  It is also worth noticing that tails of
collimated emission from diffused lobes can be quite common especially when
deep radio observations are available.  One example are the recent images of
3C449 (Feretti et al.  1998b) where a narrow structure (that appears as a
channel with edge brightening) is emerging from the inner lobe.

Norman \& Balsara (1993), following an earlier suggestion of Burns (1986),
suggested that the jet deflection and disruption in WAT radio galaxies can be
due to a jet-cloud collision in a clumpy external medium.  However, after this
the jet can be ``restarted'' even to a mildly supersonic velocity via a de
Laval nozzle mechanism (i.e.  a variation in the cross-section of the plasma
flowing channel) as shown in Fig.~4 of Norman \& Balsara (1993). 

The inner jet in Centaurus~A appears to undergo a strong oblique shock at few
arcsec from the nucleus where the jet decollimates and decelerate forming a
turbulent, subsonic flow.  At the same position it also shows a sharp bend. 
The inner lobe appears to be confined by the IGM/ISM (possibly the
same medium causing the shock) and this would explain the high brightness in
this lobe.  These characteristics bear similarities with WAT sources, as already
pointed out in the simulation of Norman et al. (1988), therefore it appears to
be possible to apply the model of Norman \& Balsara also to the case of
Centaurus~A.  For this object, the plasma accumulated in the inner lobe would
be able to ``burst'' out only through one nozzle (a sort of ``hole in the
bubble'') that would be the region where the large-scale jet forms.  The
location of the nozzle must represent either a region where the pressure
gradient is more favourable (e.g.  the edge of the cloud in the external
medium responsable of the interaction and bending) or the region where an old
jet was located if we believe the multiple outburst and change of axis
picture described above. 

In this scenario the effect of the environment in shaping the radio morphology
is really important.  The above picture requires the presence of an asymmetry
or clumpiness in the external atmosphere.  This can be justified again by the
fact that Centaurus A is believed to be a result of a recent merger and this
could have influenced the external medium as well.  

Moreover, the presence of X-ray emission in the region of the NML shows that
indeed some differences are present in this region: an asymmetry in the
external medium would indeed be consistent with the asymmetry observed in the
radio, between the northern and southern side. 

The bursting bubble model would also offer an explanation for the origin of
the inner filament (see \S 4.2)

\subsection{The optical filaments and radio plasma}

One of the aims of our observations of Centaurus~A is to investigate
whether higher resolution radio images can give additional information on the
nature and origin of the filaments of highly-ionized gas that are found
between the inner radio lobe and the NML. 

Fig.~7 shows an overlay between our higher resolution image and an optical
image kindly provided by D.~Malin.  Filament {$A$} is the so-called inner
filament, while $B$ is usually referred to as the outer filament.  The other
two filaments, {$C$} and {$D$}, represent filaments at larger radii that have
been studied by Graham \& Price (1981). 

As summarized in the introduction, from a spectroscopic study of the
filaments $A$ and {$B$}, Morganti et al.\ (1991, 1992) conclude that they are
predominantly photoionized by the radiation field of a nuclear continuum
source.  However, large velocity gradients (up to 100 \kms; Graham \& Price
1981 and Morganti et al.\ 1991), and line splitting up to 500 km s$^{-1}$ on a
scale of a few hundred parsec (Evans et al.\ in prep.) are observed in the
inner filament.  These gradients cannot be accounted for, e.g., by
acceleration of the clouds by radiation pressure of the photon beam, but {\sl
acceleration by a particle beam appears to be necessary} (Taylor, Morganti \&
Fosbury 1992).  A possible scenario is, therefore, that the filaments are
compressed and kinematically disturbed by the interaction of the radio plasma
and photoionized by the UV radiation from the nucleus.  The idea of an
interaction between the filaments and the radio jet was further investigated
by Sutherland et al.\ (1993), who found that an alternative explanation for
the ionization of the filaments could be shocks resulting from this
interaction.  In their model, the mechanical flux of a mildly supersonic
low-density jet interacting with dense clouds at the location of the filaments
is sufficient to energize the shock waves through the production of supersonic
turbulent velocities in the dense cloud via a Kelvin-Helmholtz (K-H)
instability. The energy radiated away by shocks ionizes the filaments. 

Since the location of the radio jet on the large scale was not known before, it
was difficult to judge how the interaction between the radio plasma and the
filaments was occurring, and how relevant this interaction could be for the
ionization of the filaments.  Discussions in the literature about the nature of
the filaments have always assumed that, at the location of the filaments, the
radio jet is at the same position angle as the inner jet.  However, our images
show that the radio jet is at a slightly different location and we will discuss
in the light of this result how the radio jet can influence the kinematics of
the filaments.

\begin{figure*}
\centerline{\psfig{figure=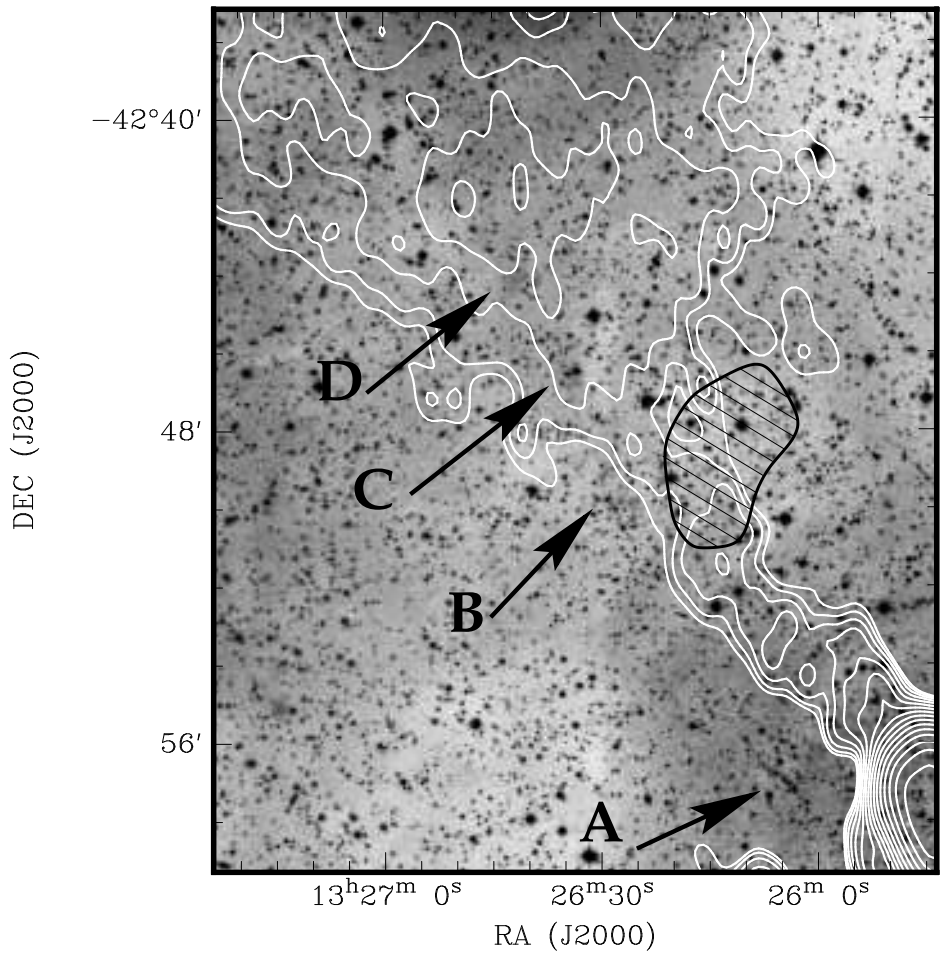,width=13cm,angle=-90}} 
\caption{Overlay between our higher resolution image and an optical
image kindly provided by D.~Malin.  The optical image has been ``unsharp
masked'' to show the optical filaments more clearly. The shaded region
corresponds to the \HI\ cloud from Schiminovich et al. (1994), see text.}
\end{figure*}

\subsubsection{The inner filament and its complex velocity field}

Fig.~7 shows that the inner filament falls about 2 arcmin ($\sim$2 kpc) away
from the large-scale radio jet revealed in our new images.  It therefore
appears that the inner filament {\sl is not directly interacting} with the
radio plasma.  The gap observed between the large-scale radio jet and the
inner filament $A$ makes it not straightforward to apply the explanation
proposed by Sutherland et al.\ (1993) for the ionization of this filament.

A few alternative hypotheses can be considered.  One possibility 
is that the radio jet is actually broader than we observe, i.e.\ we
have detected only the "spina dorsale" of the jet.  The region disturbed
by the passage of the radio jet would therefore be more extended.  However,
the lower resolution image we have available (Fig.~4) does not show a much
broader jet.  

Another possibility is that the inner filament was actually formed in the
boundary layer of the radio jet, but then the rotation of the system of which
the filaments are part carried this gas away from the jet.  The average
velocity  of the filament (V$_{\rm hel}\sim$ 280 -- 320 km s$^{-1}$) appears to
be close to the velocity of the neutral hydrogen in the eastern tip of the
dust-lane (Schiminovich et al.  1994).  This suggests that filament $A$
originally was a \HI\ cloud acquired by the merger that produced the dust
lane: the acquired gas is still not settled as suggested by the presence of a
warp in the dust lane, and the filament could represent a remnant tail of the
gas falling into the stable orbit of the dust-lane plane.  If the filament
rotates with velocities $\sim 250$-$300$ km s$^{-1}$ (as the neutral gas
appears to do), it will travel the 2 kpc that separate them from the radio jet
in at least $\sim 7 \times 10^6$ yr.  It is unlikely that the clouds would
survive such a time.  The clouds in the filament do not appear to be in
thermal equilibrium with the environment.  As shown in Morganti et al.\ (1991),
the bright blobs observed in the inner filament would reach pressure
equilibrium with the surrounding in a short time: $\leq 10^6$ yr.  Thus, it
does not seem feasible that the most compact blobs have travelled from the
edge of the radio jet to the present position and still show such strong
velocity gradients and overpressure.

Centaurus~A is not the only object that shows the presence of disturbed regions
of ionized gas at much larger radii than the radio jet.  In other objects like,
e.g., the Seyfert galaxy Mrk 3 (Capetti et al.\ 1999) and the radio galaxy
PKS~2250--41 (Villar-Martin et al.\ 1998a) similar situations have been
observed.  In the former case, Capetti et al.\ interpret this as the result of
a hot rapidly expanding cocoon surrounding the radio jet.  The hot cocoon would
be produced in the ISM/IGM heated by the interaction with the jet.  Numerical
simulations on jet instabilities and exchange of mass, momentum and energy
between the jet and the environment have been carried out by, e.g., Bodo et al. 
(1995, 1997).  In particular, in a so-called acoustic phase, the jet is able to
perturb the external medium by radiating acoustic and shock waves.  These
simulations concentrate on high Mach number ($M >$ 5) jets and therefore they
may not be fully applicable to the large-scale jet in Centaurus~A.  However, in
Centaurus A the filament lies in a ``complex'' region, close to where the
large-scale jet is actually bursting out of the inner lobe if the origin of the
large-scale jet is as described in \S 4.1.2.  It is, therefore, likely that
strong instabilities are created by these effects and propagated to the
distance of the filament.  This could create the instabilities that lead to the
large velocity gradients that is observed in this filament, in particular on
the side facing the radio jet.  The inner filament would then not be disturbed
by the jet itself, but by effects created by the jet emerging from the inner
lobe.  This hypothesis may indirectly support the idea of bursting bubble as
origin for the large-scale jet. 

\subsubsection{The outer filaments: a more direct interaction}

The outer filaments $B, C$ and $D$ appear to be, at least in projection, closer
to the radio emission and hence the case for a direct interaction between the
jet and the filaments is strong. 
It is worth remembering that for the outer filament $B$ there is a strong case
for an association between it and the \HI\ cloud (Schiminovich et al.\ 1994;
see shaded region in Fig.~7) based on the agreement between their velocities (V$_{\rm hel}
\sim$ 350 km s$^{-1}$ for the ionized gas; Graham 1998 and Evans et al. in prep). 
In this context, it is worth mentioning that new, high spectral-resolution
observations of the \HI\ cloud (Oosterloo et al.\ in prep.) show that the \HI\
closest to the filament  could be disturbed kinematically and that this side of
the \HI\ cloud has a very similar velocity gradient to that observed in the
ionized gas closest to the \HI\ cloud. 
Moreover, right at the edge of the \HI\ cloud facing the filament $B$, there
is a chain of blue stars that could point at star formation induced by the jet
(Graham 1998).

In the case of the filament $B$, most investigators have assumed in the
past (by extrapolating the position angle  of the inner radio jet) that the radio jet
passes on its eastern side.
However, using our new radio data it appears that a more complicated
interaction than has been  previously assumed has to be occuring.  Our data shows that the
jet passes on the western side of the filament, and in fact passes, in
projection,  over the brighter part of the \HI\ cloud.  Even more, the regions of ionized gas
that show a complex velocity structure or large velocity gradients (in
particular the most bright and compact regions) lie on the eastern side of the
outer filament, away from the \HI\ cloud and away from the radio jet.  The
ionized gas between these compact regions and the jet have a much more relaxed
velocity field.  It therefore appears that the source of these turbulent
velocities would come from the eastern side of the filament, while the radio
jet is on the western side of it.

Thus, if the jet is responsible for creating the outer filament by interacting
with the \HI\ cloud, one would have to assume that the \HI\ cloud is in front
or behind the jet and that the turbulent layer around the jet is quite thick
(in order to explain the distance between the jet and the compact knots in the
outer filament) and that it has quite an irregular shape (in order to explain
the offset of the knots from the \HI\ cloud). 

From the new velocity data obtained by Evans et al.\ (in  prep.) it is evident
that going  from the outer  filament $B$ (the  side facing, in projection, the
radio emission) to the filaments $C$ and $D$ the velocities go from about --200
km s$^{-1}$ with respect to the  systemic velocity to  about --1200 km s$^{-1}$
in filament $D$ (Graham \& Price (1981), Evans in  prep.).  Such very high velocities
are unlikely to have a  gravitational origin.  A  possible explanation is that
filaments $C$ and $D$  originally were  gas clouds similar  to the  \HI\ cloud
observed next to filament $B$ (at a similar location with similar velocities),
but that  they  are being carried outwards  by  a fast  outflow of   the radio
plasma.

\subsubsection{Clues on the ionization mechanism?}

In summary, our radio image does not support a simple jet-cloud interaction
picture for the filaments and a more complicated scenario is required.  It is
therefore still quite possible that both the effects of an interaction with
material powered by the jet {\sl and} photoionization from the central AGN are
needed to explain all the observed properties of the filaments.  Including a
photoionization component has some advantages.  For example, we note that the
ionized gas is only found on one side of the radio emission, well aligned with
the position angle of the inner radio jet.  Assuming that this position angle
defines also the position angle of the $UV$ radiation from the nucleus, this
could support the scenario in which clouds of gas are somehow disturbed by the
passage of the radio plasma in the ISM (as discussed above), while the main
ionization mechanism is the nuclear $UV$ that is able to ionize all the clouds
in the direction of the beam with the right density.  Moreover, the new \HI\
data of Oosterloo et al.\ suggest that the \HI\ could be disturbed
kinematically by the interaction, but it is not ionized by it.  Schiminovich et
al.\ already pointed out that the boundary between the \HI\ cloud and the
filament is very sharp and this would favour ionization by photons from the
centre as the more likely explanation for the ionization of the filament. 

It is difficult to disantangle the relative importance of the two possible  
mechanisms responsable for the ionization of the gas (photoionization from the
nucleus or from shocks) in the filaments of Centaurus A.  In the case of other
radio galaxies (e.g.\ PKS 2250-41, Villar-Martin et al.\ 1998) very detailed
and high-dispersion data have been used to resolve {\sl kinematically} the
emission from the gas which has interacted with the radio jet and the emission
from the non-shocked ambient gas.  This is a study that will have to be done
as a follow up of the work presented here.

\section{Acknowledgements}

This work is based on observations with the Australia Telescope Compact Array
(ATCA), which is operated by the CSIRO Australia Telescope National Facility.
We are grateful to the staff of the Paul Wild Observatory (Narrabri) for the  
support during the observations.  We thank David Clarke for his help in
sorting out the polarization by providing us unpublished data.  We also would
like to thank Norbert Junkes for providing an electronic copy of his Parkes
image that we have used in Fig.3 and Jane Turner for helping us with the ASCA
data.  Finally, we are indebted with Roberto Fanti, Robert Laing, Karl-Heinz
Mack and Steven Spangler for useful discussions and suggestions.

\newpage

\vspace*{1.0cm}
\parindent 0.0cm

\begin{thebibliography}{}


\bibitem{} Blanco V.M., Graham J.A., Lasker B.M., Osmer P.S.  1975,  ApJ,
198, L63

\bibitem{} Bicknell G.V. 1986, ApJ, 305, 109

\bibitem{} Bicknell G.V., de Ruiter H.R., Fanti R., Morganti R., Parma P.
 1990, ApJ, 354, 98

\bibitem{} Bodo G., Rossi P., Massaglia S., Ferrari A., Malagoli A., Rosner R.
1998, A\&A, 333, 1117

\bibitem{} Bodo G., Massaglia S., Rossi P.,  Rosner R., Ferrari A., Malagoli A.
1995, A\&A, 303, 281

\bibitem{} Bridle A.H., Perley R.A. 1984, ARAA, 22, 319

\bibitem{} Bridle A.H., Fomalont E.B., Cornwell T.J. 1981, AJ, 86, 1294

\bibitem{} Burns J.O., Feigelson E.D., Schreier E.J. 1983, ApJ, 273, 128

\bibitem{} Burns J.O. 1986, Can. Jour.Phys., 63, 373

\bibitem{} Capetti A., Axon D.J., Macchetto F.D., Marconi A., Winge C. 1999,
ApJ, in press (astro-ph/9811381)
 
\bibitem{} Chambers K.C., Miley G.K., van Breugel W. 1987,
Nature, 329, 604

\bibitem{} Clark N.E., Tadhunter C.N., Morganti R., Killeen N.E.B.,
Fosbury R.A.E., Hook R.N., Siebert J., Shaw M.A. 1997, MNRAS, 286, 558
   
\bibitem{} Clark N.E., Axon D.J., Tadhunter C.N., Robinson A., O'Brien P.
1998, ApJ, 494, 546

\bibitem{} Clarke D.A., Burns J.O., Norman M.L. 1992, ApJ, 395, 444

\bibitem{} Cooper B.F., Price R.M., Cole D.J. 1965, Aust. J. Phys., 18, 589

%\bibitem{} de Ruiter H.R., Parma P., Fanti C., Fanti R. 1990, A\&A, 227, 351

\bibitem{} Fanaroff B.L., Riley J.M. 1974, MNRAS, 167, 31p

\bibitem{} Feigelson E.D., Schreier E.J., Delvaille J.P., Giacconi R., Grindlay
J.E., Lightman A.P.  1981, ApJ, 251, 31

\bibitem{} Feretti L., Giovannini G., Klein U., Mack K.-H., Sijbring L.G., Zech
G.  1998a, A\&A, 331, 475 

\bibitem{} Feretti L., Perley R., Giovannini G., Andernach H.  1998b, A\&A, in
press (astro-ph/9810305)

\bibitem{} Fosbury R.A.E., Morganti R., Wilson W., Ekers R.D., di Serego  
Alighieri S., Tadhunter C.N. 1998, MNRAS, 296, 701

\bibitem{} Graham J.A., Price R.M. 1981, ApJ, 247, 813

\bibitem{} Graham J.A. 1998, ApJ, 502, 245

\bibitem{} Haynes R., Cannon R., Ekers R. 1983, Proc.A.S.A., 5, 241

\bibitem{} Hui X., Ford H.C., Ciardullo R., Jacoby G.H. 1993, ApJ, 414, 463

\bibitem{}Jones D.L., Tingay S.J., Murphy D.W. et al. 1996, ApJ, 466, L63

\bibitem{}Junkes N., Haynes R.F., Harnett J.I., Jauncey D.L.  1993, A\&A,
269, 29

\bibitem{} Klein U.  1998, Lecture Notes in Physics, Proceedings of a workshop
at Ringberg Castle on {\sl M\,87}, H.-P.  R\"oser and K.  Meisenheimer (eds.),
Springer, Heidelberg

\bibitem{}Killeen N.E.B., Bicknell G.V., Ekers R.D., 1988, ApJ, 325, 180

\bibitem{} Israel F.P. 1998, A\&A Rev., in press (astro-ph/9811051)

\bibitem{}Laing R.A.  1993, in ''Astrophysical Jets'', eds.  Bulgarella D.,
Livio M.  \& O'Dea, Cambridge Univ.  Press, p.95

\bibitem Mack K.-H., Kerp J., Klein U. 1997, A\&A, 324, 870

%\bibitem{}Malin D.F., Quinn P.J., Graham J.A. 1983, ApJ, 272, L5

\bibitem{}McCarthy P.J., van Breugel W., Spinrad H., Djorgovski S.  1987, ApJ,
321, L29

\bibitem{}McCarthy P.J. 1993, ARAA, 31, 639 

\bibitem{}Morganti R., Fosbury R., Hook R., Robinson A., Tsvetanov Z. 1992,
MNRAS, 256, 1p

%\bibitem{}Morganti R., Robinson A., Tsvetanov Z., Oosterloo T.  \& Fosbury R.,
%1992b,  in ``Physics of Active
%Galactic Nuclei'' Duschl W.J.  \& Wagner (eds.) Springer-Verlag

\bibitem{}Morganti R.,Robinson A., Fosbury R.A.E., di Serego Alighieri S.,
Tadhunter C., Malin D. 1991, MNRAS, 249, 91

\bibitem{}Norman M.L., Burns J.O., Sulkanen M.E. 1988, Nature, 335, 146

\bibitem{}Norman M.L.  \& Balsara D.S., 1993, in ``Jets in Extragalactic
Radio Sources'', R\"oser H.-P.  \& K.  Meisenheimer, eds., Springer-Verlag,
p. 229

%\bibitem{}Parma et al. 1998 A\&A, submitted

\bibitem{} Pacholczyk A.G. 1970, Radio Astrophysics, Freeman, San Francisco

\bibitem{} Pringle J.E. 1997, MNRAS, 292, 136

\bibitem{} Reynolds J.E. 1996, in  ``Australia Telescope Compact Array
User's Guide'', W.M. Walsh \& D.J. McKay,  eds

\bibitem{} Sault R.J., Teuben P.J., Wright M.C.H.  1995, in Astronomical
Data Analysis Software and Systems IV, R.  Shaw, H.E.  Payne and J.J.E. 
Hayes (eds), Astronomical Society of the Pacific Conference Series 77, p.  433


\bibitem{}Schiminovich D., van Gorkom J.H., van der Hulst J.M., Kasow S. 
1994, ApJ, 423, L101

\bibitem{} Schreier E.J., Marconi A., Axon D., Caon N., Macchetto D., Capetti
A., Hough J.H., Young S., Packham C. 1998, ApJ, 499, 143

\bibitem{}Sutherland D., Bicknell G., Dopita M. 1993, ApJ, 414, 510

\bibitem{}Taylor D., Morganti R.  \& Fosbury R., 1992, in  ``Extragalactic
Radio Sources -- from beams to jets'' 7th IAP Meeting Roland J., Sol H.  and
Pelletier R. (eds.) Cambridge Univ.  Press, p. 341

\bibitem{}Tingay S.J. et al. 1998, ApJ, 115, 960 

\bibitem{} van Breugel W., Miley G., Heckman T. 1984, AJ, 89, 5

\bibitem{}Villar-Martin M., Tadhunter C.N., Morganti R., Axon D., Koekemoer A. 
1998a, MNRAS, submitted

\bibitem{}Villar-Martin M., Tadhunter C.N., Morganti R., Clark N., Killeen N.,
Axon D.  1998b, A\&A, 332, 479

\bibitem{}Wright A.  \& Otrupcek R.  1990, Parkes Catalogue, Australia
Telescope National Facility

\end{thebibliography}
\end{document}